\documentstyle[12pt]{article}
\begin{document}

\def\be{\begin{equation}}
\def\ee{\end{equation}}
\def\br{\begin{eqnarray}}
\def\er{\end{eqnarray}}
\def\bc{\begin{center}}
\def\ec{\end{center}}
\def\ks {\not\!k}
\def\ps {\not\!p}
\def\qs {\not\!q}
\def\qqs {\not\!{\tilde q}}
\def\ds {\not\!\partial}
\def\eps {\not\!{\epsilon}_\lambda}
\def\dmu {\partial_{\mu}}
\def\dmmu {\partial^{\mu}}
\def\dbeta {\partial_{\beta}}
\def\dbbeta {\partial^{\beta}}
\def\dlambda{\partial_{\lambda}}
\def\dllambda{\partial^{\lambda}}
\def\dnu{\partial_{\nu}}
\def\dnnu{\partial^{\nu}}
\def\d {\partial}
\def\xs {\not\!x}
\def\ps {\not\!p}
\def\ol{\overline}
\def\piNg{\pi N \gamma}
\def\piN{\pi N}
\def\M {{{\cal M}}}
\def\G {{{\cal T}}}
\def\T {{{\cal T}}}
\def\U {{{\cal U}}}
\def\V {{{\cal V}}}
\def\Lh{\hat{\cal L}}
\def\L {{{\cal L}}}
\def\ra{\rightarrow  }
\def\qv {\vec{q}}
\def\vv {\vec{v}}
\def\qvp{\vec{q}~'}
\def\qvpp {\vec{q}~"}
\def\pv {\vec{p}}
\def\pvp{\vec{p}~'}
\def\pvpp {\vec{p}~"}
\def\xv {\vec{x}}
\def\xvp {\vec{x}~'}
\def\tv {\vec{\tau}}
\def\elmu {\epsilon_{\lambda~\mu}}
\def\elaalpha {\epsilon_{\lambda}^{~\alpha}}
\def\elmmu {\epsilon_{\lambda}^{~\mu}}
\def\el {\epsilon_{\lambda}}
\def\kv {\vec{k}}
\def\NPi#1{{1\over\sqrt{2\omega(#1)}} }  
\def\NN#1{\sqrt{{m_N \over \epsilon(#1)}} }  
\def\Nk{{1\over \sqrt{2 \omega_{\gamma}(\kv)}}}  
\def\N3{{1\over (2\pi)^3}}    
\def\rf#1{{(\ref{#1})}}
\def\rfto#1#2{{(\ref{#1}-\ref{#2})}}
\def\E#1{E(#1)}
\def\W#1{\omega(#1)}
\def\wg{\omega_{\gamma}}
\def\ket#1{|#1 \rangle}
\def\bra#1{\langle #1|}
\def\ubar#1{\overline{u}(#1)}
\def\u#1{u(#1)}
\def\Top{\hat{T}}
\def\Mop{\hat{M}}
\def\Jop{\hat{J}}
\def\Mhop{\hat{{\hspace{-.8mm}\tilde M}}}
\def\Uop{\hat{U}}
\def\Vop{\hat{V}}
\def\tVop{\hat{\tilde V}}
\def\Mop{\hat{M}}
\def\Vhop{\hat{{\hspace{-.8mm}\tilde V}}}
\def\Gop{\hat{G}}
\def\Gammaop{\hat{\Gamma}}
\def\Gamopmu{\hat{\Gamma}_{\mu}}
\def\Gamopmunu{\hat{\Gamma}_{\mu\nu}}
\def\Gamopnumu{\hat{\Gamma}_{\nu\mu}}
\def\Gamopnumualpha{\hat{\Gamma}_{\nu\mu\alpha}}
\def\Gamopmunualpha{\hat{\Gamma}_{\mu\nu\alpha}}
\def\psix{\psi(x)}
\def\psixb{\bar{\psi}(x)}
\def\pix{\vec{\pi}(x)}
\def\pixx{\pi_3(x)}
\def\psidxmu{\psi_{\Delta}_{\mu}(x)}
\def\psidxnu{\psi_{\Delta}_{\nu}(x)}
\def\psidxmmu{\psi_{\Delta}^{\mu}(x)}
\def\psidxnnu{\psi_{\Delta}^{\nu}(x)}
\def\psidxbmu{\bar{\psi}_{\Delta \mu}(x)}
\def\psidxbmmu{\bar{\psi}_{\Delta}^{\mu}(x)}
\def\psidxbnnu{\bar{\psi}_{\Delta}^{\nu}(x)}
\def\psidxbalpha{{\bar{\psi}}_{ \Delta \alpha}(x)}
\def\psidxbaalpha{{\bar{\psi}}_{ \Delta}^{\alpha}(x)}
\def\rhoxmu{\vec{\rho}_{\mu}(x)}
\def\rhoxmmu{\vec{\rho}^{~\mu}(x)}
\def\rhoxnnu{\vec{\rho}^{~\nu}(x)}
\def\g5{\gamma_5}
\def\gmu{\gamma_{\mu}}
\def\gmmu{\gamma^{\mu}}
\def\gnu{\gamma_{\nu}}
\def\gnnu{\gamma^{\nu}}
\def\galpha{\gamma_{\alpha}}
\def\gaalpha{\gamma^{\alpha}}
\def\gmmunnu{g^{\mu\nu}} 
\def\gnnummu{g^{\nu\mu}}  
\def\gmualpha{g_{\mu\alpha}} 
\def\gmmuaalpha{g^{\mu\alpha}} 
\def\gnualpha{g_{\nu\alpha}} 
\def\gnnuaalpha{g^{\nu\alpha}} 
\def\smunu{\sigma_{\mu\nu}}
\def\gmunu{g_{\mu\nu}}
\def\gnumu{g_{\nu\mu}}
\def\snnubbeta{\sigma^{\nu\beta}} 
\def\salphabeta{\sigma_{\alpha\beta}} 
\def\fpimpi{\left({f_{\pi} \over m_{\pi}}\right)} 
\def\fpiNdmpi{\left({f_{\piN\Delta} \over m_{\pi}}\right)} 
\def\grhopipi{g_{\rho\pi\pi}}
\def\endauthors{}
\def\authors#1\endauthors{#1}


\bc
{\Large\bf Dynamical model for Pion $-$ Nucleon Bremsstrahlung }
\ec
\vskip .1in

\authors
\centerline{A. Mariano$^{a,b}$ and G. L\'opez Castro$^{a}$}
\vskip .15in
\centerline{\it ${}^a$ Departamento  de F\'\i sica, Centro de
Investigaci\'on y  de Estudios Avanzados del IPN}
\centerline{\it A.P. 14$-$740 M\'exico 07000 D.F.}
\centerline{\it ${}^b$ Departamento  de F\'\i sica, Facultad de Ciencias
Ex\'actas, 
Universidad Nacional de La Plata}
\centerline{\it cc.67, 1900 La Plata, Argentina}
\endauthors

\vskip 1in

\begin{center}

{\large \bf Abstract}

\end{center}

A dynamical model based on effective Lagrangians is proposed to describe 
the bremsstrahlung reaction $ \pi N \rightarrow \pi N \gamma$ at low
energies. The $\Delta(1232)$
degrees of freedom are incorporated in a way consistent with both, 
electromagnetic gauge invariance and invariance under contact
transformations.  The model also includes the initial and final state
rescattering of hadrons via a T-matrix with off-shell
effects. The $\pi N \gamma$ differential cross sections are calculated using three different T-matrix models and the results are compared with the soft photon approximation, and with experimental data.
The aim of this analysis is to test the off-shell behavior of the different T-matrices under consideration.


\newpage

\begin{center}
{\large\bf I. INTRODUCTION}
\end{center}

 In order to extract resonant parameters of the nucleon resonances (N$^*$)
from the $\gamma N \rightarrow \pi N$ reaction, it is
important  to evaluate the background contribution to isolate the
resonant peak.
An important contribution to the background of this
photo-production reaction is provided by the final state rescattering
(FSI) of the  $\pi N$ system \cite{Nozawa1,Lee,Gross1,Sato}. Consequently,
we require knowledge of the T-matrix in the off-momentum-shell regime(off-shell) to describe this rescattering process. That is, we need information about the amplitude $T(\qvp,\qv;z(\qv))$ with
$|\qv| \ne |\qvp|$, where $z$ which is the total energy of the $\piN$ system
is a function of the relative momentum $|\qv|$ of the initial state. This
particular rescattering amplitude is more properly called the half-off-shell
T-matrix. As is known, the T-matrix can be generated by solving an integral equation of the Lippman-Schwinger or Bethe-Salpeter type by iteration of a pure phenomenological \cite{Hamada,Reid} potential, or from 
an effective potential based on meson-exchange models
\cite{Jennings,Surya}. In all
cases the so called `realistic' interactions
are fitted to reproduce the phase shifts in elastic $\piN$ scattering,
which only depends on the on-shell ($|\qv| = |\qvp|$) values of the
relative momenta.
Thus elastic scattering is not useful to constrain the
off-shell behavior of the T-matrix, i.e., interactions which yield  similar
results for elastic scattering may have different behavior in the off-shell
regime.

Another reaction where the $\piN$ off-shell T-matrix is required   is 
$ \piN \rightarrow \piNg $ bremsstrahlung. This process
has been studied \cite{Liou85, Liou88, Liou92} within the Soft-Photon Approximation (SPA).
The soft-photon amplitude, defined by the first two terms of the soft-photon expansion, depends only on the electromagnetic constants of  $\pi$ and $N$ and on the corresponding $\pi N$ elastic scattering amplitude (on-shell $\pi N$ T-matrix) \cite{Low}. It should be pointed out that Low's original amplitude fails to describe the $\pi^ \pm N \gamma$ process near the resonance region \cite{Liou85}. 
Nevertheless the SPA for the $\piNg$ reaction, as  implemented  by Ref.\cite{Liou88}, describes well the experimental data near the $\Delta$ resonance region and can provide a determination of the $\Delta^{++}$ magnetic moment \cite{Liou92}.

Because the soft-photon amplitude  depends only on the on-shell T-matrix, the information on the off-shell behavior of the T-matrix can be tested by adding the contributions to the radiative $\pi N$ scattering within the framework of a specific dynamical model.
The purpose of the present paper is to check  the off-shell
behavior of three different T-matrices for $\pi N$ rescattering in the
reaction $\pi N \rightarrow \pi N \gamma$.
A similar analysis of these FSI effects using different rescattering
off-shell amplitudes has been done with the aim to determine the form
factors of nucleons \cite{Nakayama3}.
It was found that pion photo-production reactions  are very sensitive to
the off-shell behavior of the $\piN$ interaction, and also that there are
certain inconsistencies in fixing phenomenological form factors, to match
photo-production current and FSI effects.

For this purpose, we use a dynamical model to describe the $
\piN \rightarrow \piNg $ reaction. The gauge invariant electromagnetic
current is constructed explicitly, with vertexes and propagators
 derived from the relevant hadronic and electromagnetic Lagrangians. We
also include two-body meson exchange currents, and the full
energy-momentum dependence of the $\piN$ T-matrix which exhibits its off-shell behavior. 
Finally we implement this model with different T-matrices in order to
compare their different off-shell dependencies.

 This paper is organized as follows. In section II we will construct 
the gauge-invariant amplitude for radiative $\piN$ scattering.
In section III we give a summary of the corresponding results obtained
in the SPA approximation in order to make comparisons with our dynamical
model. The Lagrangians and propagators used to
construct the gauge invariant current for our process are provided in
section IV. Finally the results and conclusions are given in section V.

\begin{center}
{\large\bf II. GAUGE INVARIANT BREMSSTRAHLUNG AMPLITUDE}
\end{center}

In the pion-nucleon bremsstrahlung process we deal with a 
problem of the scattering by two potentials \cite{Golderberger}:  
the strong pion-nucleon  and the electromagnetic interactions.
The cross section for $ \piN \rightarrow \piNg $ process reads

\br
d\sigma &=& \int {d\kv \over \wg }\int{d\qv_f \over \W{\qv_f}}
\int{d\pv_f \over \E{\pv_f}} (2\pi)^4 \delta^4(p_i+q_i -p_f-q_f-k)   
\nonumber \\ 
& & \times {1\over2}
 \sum_{\el,ms_f,ms_i}  \left| {m_N^2 \over 2\sqrt{2}}
M_{\piNg,\piN}(\el,k;q_f,p_f,ms_f;q_i,p_i,ms_i)
\right|^2,\label{1}
\er
where $q= (\omega, \qv )$, $p = (E, \pv)$ and
$k =(\omega_{\gamma}, \kv)$ denote pion, nucleon and photon
four-momenta, respectively;  $ms$ is the nucleon's spin projection and 
$\el$ indicates the polarization four-vector of the photon. The subindexes
$i,f$ refer to initial and final state quantities. 

 The Lorentz invariant amplitude\footnote{Throughout this paper, $M$
will denote the amplitude generated by the operator $\Mop$,ie.,
$M=\bra{\overline u}\Mop\ket{u}$.}  $M_{\piNg,\piN}$ explicitly reads
\br
& & M_{\piNg,\piN} = \bra{\ubar{\pv_f,ms_f}}
\Mop_{\piNg,\piN}(\el,k;q_f,p_f;q_i,p_i)\ket{\u{\pv_i,ms_i}},
\label{2'}
\er
where $\u{\pv,ms}$ denote nucleon Dirac spinors, and the amplitude
operator $\Mop_{\piNg,\piN}$ is obtained from the coupled channel
Bethe-Salpeter equation for the $\piNg$ system as follows 
(we consider electromagnetic interactions at the lowest order):
\begin{eqnarray} 
\Mop_{\piNg,\piN}  &=&   \Vop_{\piNg,\piN}\nonumber \\
&  &\hspace{-0.5cm} + i \int {dq^4 \over (2\pi)^4}
\left[\Vop_{\piNg,\piN}(q) \Gop_{\piN}(q) \Mop_{\piN,\piN}(q) +
\Mop_{\piN,\piN}(q)
\Gop_{\piN}(q) \Vop_{\piNg,\piN}(q)\ \right]\nonumber \\
&  &\hspace{-0.5cm} + i^2\int {dq^4 \over (2\pi)^4} {dq'^4 \over
(2\pi)^4}\left[ 
\Mop_{\piN, \piN}(q')\Gop_{\piN}(q')\Vop_{\piNg,\piN}(q', 
q)\Gop_{\piN}(q)\Mop_{\piN,\piN}(q)\right]
\ . \label{7}
\end{eqnarray} 
The symbol  $\int dq^4$ will indicate integration over intermediate
four-momenta variables.

In terms of the above operator amplitude, the T-matrix, defined as 
\be
\Top(q_f,p_f;q_i,p_i) = \N3 \Mop_{\piN,\piN}(q_f,p_f;q_i,p_i),\label{10}
\ee
satisfies the integral equation

\br
\Top & = & \Uop + i \int {dq^4 \over (2\pi)^4}
 \Uop(q) \Gop(q) \Top(q)\label{11},\\
\Uop & = & \N3 \Vop_{\piN,\piN},\nonumber \\
\Gop & = &(2\pi)^3\Gop_{\piN}.\nonumber
\er
In the previous equations $\Vop_{ij}$ denote $\Mop$-matrix elements
corresponding to the irreducible
Feynman diagrams for each process, while $\Gop_i$ is the product of
Feynman propagators of intermediate particles.

Following the Thompson's prescription \cite{Thompson}, we can represent the above integrals in a three-dimensional form as follows (we set in the center of mass frame of the $\piN$ system)
\be
\Top(\qvp,\qv,z) = \Uop(\qvp,\qv) + \int d^3 \qv'' \Uop(\qvp,\qv'') 
\Gop_{TH}(z,\qv'') \Top(\qv'',\qv,z),\label{13}
\ee
with,
\be
\Gop_{TH}(z,\qv'') = {m_N \over 2 \W{\qv''} \E{-\qv''}}
{\sum\limits_{ms''}\ket{\u{-\qv'',ms''}}\bra{\ubar{-\qv'',ms''}}\over
z - z'' + i\eta},\label{14}
\ee
where $z'' = \E{-\qv''}+\W{\qv''}$. In the above expressions $\Gop_{TH}$
denotes the Thompson
propagator replacing the full $\Gop_{\piN}$~ Feynman propagator which, as
a consequence of the three-dimensional reduction, eliminates the
propagation of antiparticles and puts intermediate particles on their 
mass-shell. 
The kernel function $\Uop(\qvp,\qv)$ contains all the $\piN$-interaction
irreducible
diagrams to be iterated in the T-matrix calculation, but usually only
second-order contributions are kept.

The electromagnetic current  $\Vop_{\piNg,\piN}$ can be broken into two
pieces,

\be
\Vop_{\piNg,\piN} \equiv \Vop_{\piNg,\piN}^{(1)} +
\Vop_{\piNg,\piN}^{(2)},\label{17}
\ee
where the upper indices denote one- and two-body contributions,
respectively, which are obtained by coupling the photon to all the internal
lines of  $\Uop$. As is known, the operator $\Vop_{\piNg,\piN}^{(2)}$ must
be added to the electromagnetic current in order to satisfy the 
electromagnetic gauge invariance of the total amplitude \cite{Nakayama1},
while the one-body amplitude $V_{\piNg,\piN}^{(1)}$ vanishes for free
hadrons. Both contributions to the  total amplitude are illustrated in 
Fig. 1.

Let us now discuss some problems related to gauge invariance.
 The bremsstrahlung amplitude can be directly computed 
from Eqs.\rfto{2'}{10}, by making a reduction that replaces $\Gop_{\piN}$
by $\Gop_{TH}$ in Eq.\rf{7} \cite{Mariano}. This procedure, however, 
introduces some inconveniences.
First, $V_{\piNg,\piN}$ is not gauge invariant by itself because it
involves the addition of zero and second order terms in the hadronic
vertexes, and  thus $M_{\piNg,\piN}$ is not manifestly gauge invariant.
Second, the three-dimensional reduction destroys any possibility
of obtaining gauge invariance, because we loose the propagation of
antiparticles\footnote{As an example let us consider the amplitude for 
photon  emission off a charged pion. This contribution involves the product of the
electromagnetic vertex $\Gamma_{\pi}$ and the pion propagator
$\Delta_{\pi}$ in the following form: ${\hat
\Gamma}_{\pi}( q \pm k, q).\el^* \Delta_\pi(q\pm k) = \pm e ~(2q \pm k) .\el^*
{1 \over (q \pm k)^2 - m_\pi^2}$. When
we replace $\epsilon_{\lambda} \rightarrow k$ to verify gauge-invariance
we obtain, $\hat{\Gamma}_{\pi}(q \pm k,q).\el^*  \Delta_\pi(q \pm k) = \pm e$.
 The three dimensional reduction replaces the full propagator
$\Delta_\pi(q)$  by  $\Delta_\pi^+(q) = {1 \over 2\W{\qv}}{1\over (q^0 -
\W{\qv})}$ and such a relation is no longer fulfilled.}.

We can follow an alternative procedure that makes the bremsstrahlung
amplitude manifestly gauge invariant and where the Thompson reduction
does not introduce the problems mentioned above. If we substitute Eqs.
\rf{10} and \rf{11} into Eq.\rf{7}, only in the
one-body component of the amplitude $M_{\piNg,\piN}^{(1)}
\equiv M_{\piNg,\piN}(\Vop_{\piNg,\piN}^{(1)})$, we can isolate the
lowest order nonzero contribution of the one-body current. After the 
three dimensional reduction, the total amplitude can be rewritten as

\be
M_{\piNg,\piN} \equiv \left[ \tilde V_{\piNg,\piN} + \tilde
M^{pre}_{\piNg,\piN}  + \tilde M^{post}_{\piNg,\piN} + \tilde
M^{double}_{\piNg,\piN}\right],\label{20}
\ee
with 
\br
\tilde V_{\piNg,\piN} &  = &
\bra{\ubar{-\qvp-\kv/2,ms_f}}\tVop_{\piNg,\piN}(\el,\kv,\qvp,\qv)
\ket{\u{-\qv,ms_i}},\nonumber\\
& & \nonumber\\
\tilde M^{pre}_{\piNg,\piN}  & = &
\int dq''^3 \bra{\ubar{-\qvp,ms_f}}\Top^{(-)\dagger}(\qvp,\qv'',z')
\Gop_{TH}(z',\qv'') \tVop_{\piNg,\piN}(\el,\vec k,\qv'',\qv)
\ket{\u{-\qv+\vec{ k} /2},ms_i}
,\nonumber \\
& & \nonumber\\
\tilde M^{post}_{\piNg,\piN} & = & 
\int dq''^3
\bra{\ubar{-\qvp-\kv/2,ms_f}}\tVop_{\piNg,\piN}(\el,\kv,\qvp,\qv'')
\Gop_{TH}(z,\qv'') \Top(\qv'',\qv,z)\ket{\u{-\qv,ms_i}}
,\nonumber\\
& & \nonumber\\
\tilde M^{double}_{\piNg,\piN}& = & \int dq''^3\int dq'''^3
\bra{\ubar{-\qvp-\kv/2,ms_f}}\Top^{(-)\dagger}(-\qvp-\kv/2,\qv'',z')
\nonumber \\
& &\Gop_{TH}(z',\qv'')\tVop_{\piNg,\piN} (\el,\kv,\qv'',\qv''')
\Gop_{TH}(z,\qv''')\Top(\qv''',\qv,z)
\ket{\u{-\qv,ms_i}}\label{21}
\er
where the current
\br
\tVop_{\piNg,\piN} = i \Vop_{\piNg,\piN}^{(1)} \Gop \Uop +  i \Uop^\dagger \Gop \Vop_{\piNg,\piN}^{(1)} + \Vop_{\piNg,\piN}^{(2)},\label{22}
\er
generates  a gauge invariant Born  amplitude $\tilde V_{\piNg,\piN}$,
that involves all possible ways of attaching a photon to the $\piN$
scattering amplitude $U$, and contains the full propagator 
$\Gop \sim \Gop_{\pi N}$. The operator $\Top^{(-)}(z')$, where $z'= z +
\wg$, obeys Eq.\rf{13} if we change $\eta \rightarrow -\eta$ in Eq.\rf{14}.

The superscript  {\it pre (post)} in Eq. (9) indicates that the photon is
emitted before (after) the action of the T-matrix, while the  superscript
{\it double} refers to a double-scattering term where the photon is 
emitted from internal lines between two T-matrices. In the above
equations, the Born, pre and double amplitudes were evaluated
in  the initial center of mass frame ($\qv_f = \qvp -\kv/2,\pv_f
=-\qvp -\kv/2 $, $\qv_i = - \pv_i = \qv$),
while the other (post) amplitude was evaluated in
the corresponding final frame ($\qv_f = - \pv_f = \qvp$, $\qv_i =
\qv + \vec{k} /2,\pv_i = -\qv + \vec{k} /2$). The different terms
in Eq. (9) are illustrated in Fig. 2a ($\tilde V_{\piNg,\piN}$) and in  
Fig. 2b (remaining terms).

\bc
{\large \bf III.SOFT PHOTON APPROXIMATION}
\ec

In this section we present a brief review of the soft-photon approximation
(SPA)  to the radiative $\piN$ scattering process.
This will be helpful for the introduction of the notation and for later comparison
with our dynamical model. Within the SPA \cite{Liou85} the  T-matrices
can be represented as an expansion in powers of the photon energy
$\omega_{\gamma}$. Using
the TETA (two energy - two angle) kinematics \cite{Liou88} we can write
this expansion as follows:
\br
\Top(s,t,\Delta) & = & \Top(s,t,m^2) + {\d \Top \over \d \Delta} \times
{\d \Delta \over \d \wg}~ \wg+ O(\wg^2), \label{24}
\er
where $(s,t,\Delta)$  indicates one of the most  convenient set of
variables among $(s_i,t_p,\Delta_{q_i})$, $(s_i,t_q,\Delta_{p_i})$,
$(s_f,t_p,\Delta_{q_f})$, and $(s_f,t_q,\Delta_{p_f})$ to describe the
process.  These Lorentz-invariant variables are defined as
\br
& &\hspace{-1.cm}  s_i = (q_i + p_i)^2,~~ s_f  =  (q_f + p_f)^2,~~t_p =
(p_i - p_f)^2,~~t_q  =  (q_i - q_f)^2,\nonumber \\
& &\hspace{-1.cm} \Delta_{q_i}  =  (q_i  -  k )^2,~~\Delta_{q_f}  = (q_f
+  k )^2,~~
\Delta_{p_i}  =  (p_i  -  k )^2,~~\Delta_{p_f}  =  (p_f  +  k )^2.
\label{25}
\er
If we set $k=0$ in the previous equations, 
we get $\Delta_{q_i} = \Delta_{q_f} = m_{\pi}^2$ and 
$\Delta_{p_i} = \Delta_{p_f} = m_N^2$, {\it i.e.} they reduce to the
particle's masses in every case. Therefore, when $k \ne 0$, the $\Delta$
variables provide a convenient set to measure the off-shell character
of the intermediate particles, and the derivatives in Eq.\rf{24} account
for off-shell effects in the T-matrix.

Within the SPA, the total bremsstrahlung amplitude can be split into
external ($E$) and internal ($I$) contributions:
\br
M_{\piNg,\piN} \equiv M^{E}_{\piNg,\piN} + M^{I}_{\piNg,\piN},
\label{26}
\er
where we can identify 
\br
M^{E}_{\piNg,\piN} & \equiv &
\tilde M^{pre} _{\piNg,\piN}(\tVop_{\piNg,\piN} \rightarrow
\Vop_{\piNg,\piN}^{(1)}) + 
\tilde M^{post}_{\piNg,\piN}(\tVop_{\piNg,\piN} \rightarrow
\Vop_{\piNg,\piN}^{(1)}), \nonumber
\er
and the internal contribution $M_{\piNg,\piN}^{I}$  can be obtained ``by
imposing'' the gauge invariance condition
\be
M_{\piNg,\piN}(\elmmu = k^{\mu}) = 0. \label{28}
\ee
In this approach, the total amplitude depends only on the static
electromagnetic properties of the external particles and on the elastic
$\piN$ scattering amplitude (see Eq. (16) below) \cite{Low}. Therefore,
the total amplitude at this order eliminates any dependence on off-shell
effects and model-dependent contributions.

Up to terms of $O(\wg^0)$, the total amplitude is given by:
\br
& &~~~~~~~~~~~~~~~~~~~~~~~~~ M_{\piNg,\piN} \approx \nonumber \\
&  & \bra{\ubar{\pv'_+,m_{s_f}}}
\left\{\hat{e}_{\pi}~ \left[{ q'_{-} \cdot \el 
\over q'_{-} \cdot k} - { (q'_{-}+ p'_+) \cdot \el \over 
k \cdot (q'_{-}+ p'_+)}\right]\Top(s_i,t_p,m_{\pi}^2)\right. \nonumber \\ 
& & \left. \hspace{-1cm}
+ \left[{(\hat{e}_N~ p'_+ - \hat{R}(p'_+))\cdot \el \over p'_+ \cdot k} -
{( \hat{e}_N ~(q'_{-}+ p'_+) - \hat{R}(p'_+))\cdot \el \over k \cdot (q'_{-}+ p'_+)}
\right]\Top(s_i,t_q,m_N^2)\right\} \ket{\u{\pv,m_{s_i}}} \nonumber \\
&  & -\bra{\ubar{\pv',m_{s_f}}}\left\{\hat{e}_{\pi}~\Top(s_f,t'_p,m_{\pi}^2) \left[{q_+ \cdot \el   
 \over q_+ \cdot k} - {(q_+ + p_{-})\cdot \el \over k \cdot 
 (q_+ + p_{-})}\right]  \right. \nonumber \\ 
&  & \hspace{-1cm}\left. + \Top(s_f,t'_q,m_N^2) \left[{(\hat{e}_N ~p_{-} - \hat{R}(p_{-}))\cdot \el
\over p_{-} \cdot k} - {(\hat{e}_N~(q_+ + p_{-})- \hat{R}(p_{-}))\cdot \el
 \over k \cdot (q_+ + p_{-})}
\right]\right\}\ket{\u{\pv_{-},m_{s_i}}},
\nonumber\\
& & \label{29}
\er
where
\br
& &\hspace{-2cm}q  \equiv  (\W{\qv},\qv),~~p  \equiv  (\E{-\qv},-\qv),
~~q'  \equiv  (\W{\qvp},\qvp),~~p'  \equiv  (\E{-\qvp},-\qvp),\nonumber 
\er
\br
& & q'_{-}  \equiv (\W{\qvp_{-}},\qvp_{-}),~~p'_+\equiv(\E{\qvp_+},-\qvp_+),~~
q_+ \equiv (\W{\qv_+},\qv_+),~~p_{-} \equiv (\E{\qv_{-}},-\qv_{-}),~~
\nonumber
\er
\br
& &\hspace{-3.5cm}t_p = (p'_+ - p)^2,~~t_q = (q'_+ - q)^2,~~t'_p = (p_- -
p')^2,~~t'_q = (q_+ - q')^2,\nonumber
\er
\br
& &\hspace{-3.5cm} \hat{R}_\mu(x) \equiv {1\over 4} \hat{e}_N[\ks,\gamma_\mu] + {\hat{\kappa}_N\over 8 m_N}
\{ [\ks,\gamma_\mu],\xs\},~~
\qv_{\pm} =  \qv \pm \kv/2,~~ \qvp_{\pm} = \qvp \pm \kv/2\ ,\label{30}
\er
and $\hat{e}_{\pi}= e \T_z $, $\hat{e}_N= e(1 + \tau_z)/2$, 
and $\hat{\kappa}_N = \kappa_p (1 + \tau_z)/2 + \kappa_n (1 - \tau_z)/2$
denote the charge and anomalous magnetic moment operators of pions and
nucleons. As anticipated, Eq.\rf{29} shows that within the SPA the
$M_{\piNg,\piN}$ amplitude  depends only on the elastic T-matrix, because 
 derivative terms of $\Top$ cancels in the addition of internal and
external contributions. Let us emphasize that any dependence on the
structure of internal contributions (in particular, the dependence of
off-shell effects) are of higher order in $\wg$ and must
be included explicitly in the amplitude in a gauge invariant way. This is
the purpose of the forthcoming section.
\newpage
\begin{center}
{\large\bf IV. DYNAMICAL MODEL}
\end{center}

In this section we compute the bremsstrahlung amplitude along the lines
developed in Eqs. (9-11), using a potential $\Uop$ obtained
from effective  Lagrangians\cite{Zumino}, and three specific models for 
the T-matrix to describe the $\piN$ rescattering.

  The three models for the T-matrix will be called OBQA, SEP and
NEW, respectively.  The OBQA version for the T-matrix
interaction\cite{Schutz}, is based on a model that
includes $\pi$ and $\rho$ mesons exchange through a correlated $2\pi$
exchange potential. The SEP model for $\piN$ rescattering is generated 
\cite{Nozawa} by a  phenomenological separable potential. Finally, 
the NEW model \cite{New} is obtained from the exchange of $\pi$ and
(sharp) $\rho$  mesons. 

The operator $\Uop$ is constructed from a Lagrangian density
that includes  the nucleon ($N$), the $\Delta$-isobar, and the $\pi$, $\rho$ and $\sigma$ mesons in the following form

\be
\Lh_{hadr} = \Lh_{\pi NN} + \Lh_{N\Delta\pi} + \Lh_{\rho NN} + \Lh_{\rho
\pi\pi} + \Lh_{\sigma NN} + \Lh_{\sigma \pi\pi}
\label{31}
\ee
where the individual terms are given by
\br
\Lh_{\pi NN}(x) & = & -\left({f_{\pi N N} \over m_{\pi}}\right) \psixb \g5 
\tv.(\ds\pix) \psix , 
\nonumber 
\\
\Lh_{N\Delta\pi}(x) & = & \left({f_{N\Delta\pi} \over m_{\pi}}\right)
\psidxbmmu \vec{T}^{~\dag}.(\dmu \pix)\psix + h.c. , \nonumber 
\\
\Lh_{\rho NN}(x)& = &-{1 \over 2}g_{\rho}\psixb  \left[ \gmu - 
{\kappa_{\rho} \over 2 m_N}
\smunu \dnnu\right]\tv.\rhoxmmu\psix, \nonumber
\\
\Lh_{\rho \pi\pi}(x)& = & - \grhopipi  \rhoxmu.(\pix \wedge \dmmu
\pix),\nonumber
\\
\Lh_{\sigma NN }(x)& = & g_{\sigma}\psixb \psix \sigma(x),\nonumber
\\
\Lh_{\sigma \pi\pi}(x)& = & \left({g_{\sigma \pi \pi} \over 2
m_\pi}\right)\sigma(x)(\dmu \pix) .(\dmmu \pix). \label{32}
\er
The isotopic fields $\psix $ and $\psidxmmu$ denote the $N$ and $\Delta$
baryons, respectively, while $\pix$, $\rhoxmmu$ and $\sigma(x)$ denote
the pion, $\rho$-meson  and $\sigma$-meson fields. 
The arrow over the meson fields refers to the isospin space.
$\vec{T}^{~\dag}$
stands for the isospin $1/2$ to $3/2$ transition operator, and $f_{\pi N
N}$, $g_{\rho}$( $\kappa_{\rho}$), $g_\sigma$, $f_{N \Delta \pi}$,
$\grhopipi$,  and $g_{\sigma \pi \pi}$ are the corresponding 
coupling constants. 

The electromagnetic currents can be obtained from the Lagrangian density

\be
\Lh_{elec} = \Lh_{\gamma NN} + \Lh_{\gamma \pi NN} + \Lh_{\gamma\pi\pi} 
   + \Lh_{\gamma \Delta\Delta} + \Lh_{\gamma \pi N \Delta}
+ \Lh_{\gamma\rho\pi\pi} + \Lh_{\gamma\sigma\pi\pi}
\label{33}
\ee
with
\br
\Lh_{\gamma NN}(x) & = & - e \psixb \left[\hat{e}_N\gmu - 
{\hat{\kappa} \over 2 m_N} \smunu \dnnu \right ]  A^{\mu}(x) \psix
,\nonumber \\
\Lh_{\gamma\pi NN}(x) & = & - e \left({f_{\pi N N} \over m_{\pi}}\right) 
\psixb \g5 \gmu \left[ \tv \times \pix \right ] _3 \psix A^{\mu}(x),
\nonumber \\
\Lh_{\gamma \pi\pi}(x) & = &  - e\left[\pix \times \dmu \pix  \right]_3
A^{\mu}(x), \nonumber \\
\Lh_{\gamma \Delta\Delta}(x) & = & - e \psidxbnnu 
\Gamopnumualpha \psidxmmu A^{\alpha}(x)
,\nonumber \\
\Lh_{\gamma \pi N \Delta}(x) & = &
e\left({f_{N\Delta\pi} \over m_{\pi}}\right)
\psidxbmmu  \left [\pix \times \vec{T}^{~\dag}\right]_3 \psix A_{\mu}(x)+ h.c.,
 \nonumber 
\\
\Lh_{\gamma\rho\pi\pi}(x) & = & e g_\rho \left\{[\pix.\pix - \pixx \pixx]
\rho^\nu_3(x)\right.\nonumber \\
& &\left. - \pixx[\pix.\rhoxnnu - \pixx \rho^\nu_3(x)]\right\}A_{\nu}(x),
\nonumber \\
\Lh_{\gamma \sigma \pi\pi}(x) & = &  - 2e
\left({g_{\sigma \pi \pi} \over 2 m_\pi}\right)\sigma(x) \left[  \pix \times \dmu \pix\right]_3
A^{\mu}(x). 
\label{34}
\er
The electromagnetic vertex operator of the $\Delta$-isobar is
given by \cite{Rarita,Nath}
\br
\Gamopnumualpha & = & {\hat e}_\Delta
\left[ (\galpha
\gnumu - {1 \over 3} \galpha \gnu \gmu - {1 \over 3} \gnu\gmualpha +
 {1 \over 3} \gmu \gnualpha)
- {\hat{\kappa}_{\Delta} \over 2 m_\Delta} \salphabeta k_\beta \gnumu
\right],\nonumber \\
& & \label{35}
\er 
where $F_{\mu\nu} = \partial_{\mu} A_{\nu}(x) - \dnu A_{\mu}(x)$, $A_\nu(x)$ being the electromagnetic four-potential.
$\hat{\kappa}_{\Delta}$ and ${\hat e}_\Delta$ are the anomalous magnetic
moment\footnote{We restrict ourselves to the $\Delta^{++}$ contribution,
the only one for which we have experimental information on
$\kappa_\Delta$ \cite{Liou92}.} 
and charge operators whose action upon the Rarita-Schwinger field
give as eigenvalues  the corresponding values of these properties of the
$\Delta$-isobar.

The propagators for the $\pi (\sigma),\ \rho , \ N$ and $\Delta$
hadrons obtained from the above Lagrangians are given, respectively, by

\begin{eqnarray}
\Delta_{\pi,\sigma}(q)  &=&  {1 \over q^2 -m_{\pi,\sigma}^2 +
i\eta},\nonumber \\
 D_\rho^{\mu\nu}(q) &=& {\gmmunnu - q^\mu q^\nu /m_\rho^2  \over q^2 -
m_\rho^2 + i\eta},\nonumber \\
S(q) &=&  {\qs + m_N \over q^2 -m_N^2 + i\eta}\nonumber \\
G^{\mu\nu}(q)  &=&  {\qs + m_\Delta \over q^2 -m_\Delta^2 + i\eta}
\left[-\gmmunnu + \frac{1}{3} \gmmu \gnnu + \frac{2}{3} {q^\mu q^\nu \over
m_\Delta ^2 }- \frac{1}{3} {q^\mu \gnnu - q^\nu \gmmu \over  m_\Delta
}\right]\nonumber \\ 
& &~~~ -{
2\over 3 m_\Delta ^2}(q^2 - m_\Delta ^2)\left[{ \over }(\gmmu  q^\nu -
\gnnu  q^\mu) + (\qs + m_\Delta)\gmmu \gnnu\right]
\label{36}
\er
 Observe that we have kept the off-shell part of $G^{\mu\nu}(q)$ in Eqs.
\rf{36};
without this term it is impossible to get gauge invariance
consistently using simultaneously  the vertex given in
Eq.\rf{35} \cite{Castro}.
As was mentioned previously, the potential operator $\Uop$ can be
computed by using these Feynman rules. The  scattering amplitude $U$ is
depicted in Fig.3, while the amplitude $\tilde{V}_{\piNg,\piN}$
can be obtained by coupling the photon to all diagrams in $U$ as shown in
fig.4.
 
Good convergence properties of the scattering equations given in
Eqs. \rf{21} can be obtained by introducing hadronic form factors,
which are supposed to describe the composite nature of hadrons.
It is a common practice to use different parametrizations of the form
factors for different T-matrices.  For example, the OBQA model uses
monopole and dipole forms with cutoff parameters ranging from $\Lambda =
1200-1600 MeV$ \cite{Schutz}.
In the case of the SEP interaction different form factors are introduced for
each partial wave component \cite{Nozawa}, while in the NEW
model form factors usually advocated are of monopolar form with $\Lambda =
1300-2300 MeV$ \cite{New}.
However, the introduction of form factors replacing point vertexes in 
$\tilde V_{\piNg,\piN}$ spoils the gauge invariance of the total
amplitude. Fortunately, the gauge invariance of the amplitude can be
recovered by using the method of Gross and Riska\cite{Gross} which,
however,  does not yield  unique electromagnetic couplings to hadrons.
Therefore, we follow the more simple prescription of using a common form 
factor\cite{Nakayama3,Nozawa} of monopole type 

\be
f(\vec q'') = { \Lambda^2 \over \Lambda^2 + \vec q''^2},\label{37}
\ee
where the scale $\Lambda$  can be adjusted at a given incident energy for
each T-matrix model.

\

\begin{center}
{\large\bf V. NUMERICAL RESULTS AND CONCLUSIONS}
\end{center}

The differential cross section $d\sigma/ d\Omega_\pi d\Omega_\gamma
d\omega_\gamma$ to be compared with experimental data can be obtained from
Eq.\rf{1}, where the amplitude  $M_{\piNg,\piN}$ is calculated from
Eqs.(9-11). The dynamical model approximation (DMA) advocated in the
present paper contains the following steps. The current operator
$\tVop_{\piNg,\piN}$ is computed from the effective Lagrangians given in 
Eqs.\rf{31} and \rf{33}, with  the propagators obtained in  Eqs.\rf{36},
and the monopole form factor given in Eq.\rf{37} to have good convergence  
of the intermediate momentum integrals.

As is known, double scattering terms have a significant contribution in the
case of $proton-proton$ bremsstrahlung, mainly in the end point region of
 $\omega_\gamma$ \cite{Nakayama2}. In the present work we will neglect  
$\tilde M^{double}_{\piNg,\piN}$ in Eq.\rf{21}, because the numerical
calculation of  the three-dimensional integrals requires an enormous
computational effort. 
Nevertheless, we  keep double scattering-like contributions in  {\it post} 
and {\it pre} amplitudes coming from current components $ i \Uop^\dagger
\Gop \Vop_{\piNg,\piN}^{(1)}$ and $i \Vop_{\piNg,\piN}^{(1)} \Gop \Uop$, 
respectively. In order to compare the approaches provided by the DMA and
SPA, we fix $M_{\piNg,\piN}$ to coincide quantitatively at low
photon energies.

 For illustration purposes, we will implement  the DMA approach with
the OBQA, SEP and NEW T-matrices  for the specific example of $\pi^+ p
\rightarrow \pi^+ p \gamma$. 
In this case, we have contributions coming from the diagrams depicted
in Figs. 4a (with intermediate $\Delta$), 4b (with intermediate $N$), and
4c.  The different coupling constants and  masses needed to
evaluate $\Vop_{\piNg,\piN}$ were taken from the model II of
ref.\cite{Schutz}, from \cite{Nozawa}, and  from \cite{New}, and are
displayed in table 1.
For direct pole diagrams we use bare masses and coupling constants
since they get  dynamically dressed  in the T-matrix scattering Eq.\rf{11}.
In the OBQA case, we replace the 
$2\pi$ correlated exchange potential by $\pi$ and $\rho$  sharp mass
exchange terms, since they do not lead to sizable
differences as shown in ref.\cite{Schutz}. In the SEP case we use the same  
coupling constants and masses, because the scattering potential is not
generated from a dynamical model.
 
We will compare the theoretical predictions to the experimental cross
sections measured by  Nefkens \cite{Nefkens}(EXP I) and Meyer \cite{Meyer}  (EXP II), which have been reported for different kinematical
configurations. In EXP I the pions were detected at fixed angles 
$\theta_\pi = 50.5^0,  \phi_\pi = 180^0$, for three different energies of  
incident pions (269, 298 and 324 MeV), 
and the photons were detected at various $\theta_\gamma, \phi_\gamma$
angles in the range of energies  $\wg =0 - 150$ MeV. In EXP II  
$\theta_\pi$ ranges from 55$^0$  to 95$^0$, the
incident energy of the pion is fixed at 299 MeV and the photons were 
detected at angles  $\theta_\gamma = 120^0 , \phi_\gamma = 0^0$ with
energies in the range  $\wg =0-140$ MeV. In EXP II the resulting cross
section measurements were averaged over $\theta_\pi$.

Our results for the cross sections in  the DMA approach,  for the
respective ansatz of the T-matrix, are shown in Figs. 5 and 6.
The SPA and experimental values are also plotted for comparison. 
The predictions of the DMA approach shown in fig. 5 using the three models
of the T-matrix, are compared to the results of EXP I for photon
angles given by $G_{14} \equiv (\theta_\gamma = 103^0, \phi_\gamma =
180^0)$. Since the parameters entering  T-matrices are usually quoted to
reproduce the elastic phase shifts, the cutoff parameters $\Lambda$ were
fixed in order to have good coincidence between the predictions
of the different interactions and the SPA at low $\wg$ values. We get (in
MeV units)  $\Lambda_{OBQA} = 750,700,600$, $\Lambda_{SEP} = 700,600,500$,
and $\Lambda_{NEW} = 550,500,450$, for incident pion energies $T_{lab} =
269, 298, 324$ respectively.
Observe that the value of $\Lambda$ found in the case of the SEP
interaction for $T_{lab} = 298 MeV$ is consistent with the one previously 
found for pion photo-production at $T_{lab} = 300 MeV$\cite{Nozawa}, while
the OBQA values are roughly consistent with the form factors used in
ref.\cite{Schutz}, which correspond to a monopole form-factor with
$\Lambda \approx 800 MeV$\cite{Nakayama3}. On the other hand, 
the resulting cross sections for  the conditions of EXP II are shown  in
Fig. 6. In this case, the differential cross section is averaged over  
$\theta_\pi = 55^0 - 75^0$, and   $\theta_\pi = 75^0 - 95^0$, for $T_{lab}
= 299 MeV$. However, we keep for consistency the values of $\Lambda$
obtained at  $T_{lab} = 298 MeV$ in EXP I.

As we can observe, the SPA reproduces very well the experimental  cross
section for EXP I in the whole range of measured photon energies, while
it gives values somewhat below  those obtained in EXP II. According to
ref. \cite{Liou92}, the agreement
with EXP II might be improved if the emission of the photon from the
$\Delta^{++}$( seventh graph in fig.3a) is included explicitly  as a piece of the internal amplitude.

In almost all the cases, the predictions of the DMA  lies above the  
experimental cross section and the SPA for energies $\wg > 20 MeV$.
One of the reasons for this may be the use of an overall form factor to
cure the gauge invariance problems. The total bremsstrahlung amplitude is
built up, as can be seen from Eqs. \rf{20} and \rf{21}, by adding
different components. It is not expected that the common form factor
works as satisfactorily upon adding up these components as it does the one 
used to generate the individual T-matrices, which change their values  
from vertex to vertex.
The comparison between the results of the SPA and DMA schemes shows that
the additional off-shell effects, added coherently to the lowest order
contributions, may have important contributions since they do not cancel
exactly as the derivative terms of the T-matrix appearing from the
soft-photon expansion.
 
From fig.5 we can check that the SEP interaction provides the closest
results to the experimental cross section with deviations starting for
$\wg > 40 MeV$.  This indicates
that the dynamical model involved in the SEP interaction gives the
smallest off-shell effects. On the other hand, the strongest off-shell
effects appear in the OBQA model. This conclusion agrees with a previous
study\cite{Nakayama3} on observables in pion photo-production experiments.
Similar conclusions can be drawn from Fig. 6. Note, however, that the
SPA lies somewhat below the experimental
results in this case, which seems to indicate  the necessity of additional
dynamical degrees of freedom.

As was discussed in section III, the off-shell contributions to the
external and internal amplitudes within the SPA cancel each other, thus
we cannot study these effects within this approximation. In addition since
we get gauge invariance in the SPA by adjusting the internal amplitude,
the  gauge-invariant electromagnetic currents remain hidden.
In the DMA approach, these cancellations must occur explicitly between 
the different components of the amplitude (the so-called Born,pre,post
contributions). The departure
of the different T-matrices from the SPA can be used to estimate the size
of unbalanced off-shell terms and provide a test  of their off-shell
behaviors. Also, since the  gauge-invariant electromagnetic current is
constructed explicitly from effective Lagrangians, we can use
the radiative $\piN$  reaction to study the relevance of the degrees of
freedom and the parameters involved in this dynamical model. 

Finally one may wonder about the relevance of additional contributions
not included in the present formulation of effective Lagrangians, for
example the $a_1$, $\omega$ mesons.
A preliminary estimation of these effects shows that, in the region of
photon energies considered in this work  $\wg =0-150 MeV$, their
contribution to the amplitude are suppressed by an order of
magnitude with respect to the degrees of freedom considered here.
In addition it may be required to include the double-scattering terms, in
order to provide a better fit of experimental data. These
considerations, together with a partial wave analysis of
the studied T-matrices  and a more detailed analysis of  each particular
contribution  to the electromagnetic current are beyond the scope of the
present paper.

\begin{center}
{\large\bf ACKNOWLEDGMENTS}
\end{center}

\noindent

The work of A. Mariano was supported in part by Conacyt (M\'exico) through
the Fondo de  C\'atedras Patrimoniales de Excelencia Nivel II, and Conicet (Argentina).  He is also grateful to K. Nakayama for the valuable support in making this work.
G. L\'opez Castro was partially supported by Conacyt under contract 32429-E.

\newpage
\begin{table}[h]
\begin{center}
\caption { Relevant masses (in MeV) and coupling constants used in the
computation of $\tilde V_{\piNg,\piN}$}
\label{tab3}
\bigskip
\begin{tabular}{cccc}
\hline
&$OBQA$&$SEP$&$NEW$\\
\hline\hline
$m_{N}$ &$938.926$&$938.926$&$938.926$\\
$m_{0\Delta}$ &$1515$&$1450$&$1405$\\
$m_{\rho}$ &$769$&$769$&$769$\\
$m_{\sigma}$ &$650$&$650$&$650$\\
$f^2_{NN \pi}/4\pi$ &$0.0778$ &$0.0778$ &$0.778$\\
$f^2_{0 N \Delta \pi}/4\pi $ &$0.21$ &$0.64$ &$0.18$\\
$k_\Delta$ &$1.6$ &$1.6$&$1.6$\\
$g^2_{NN\rho}/4\pi$ &$5.05$ &$0.563$ &$0.90$\\
$k_\rho$ &$2.69$ &$3.7$&$6.1$\\
$g^2_{NN\sigma}/4\pi$&$8.94$&$8.94$&$13.$\\
$g^2_{\pi\pi\rho}/4\pi$&$5.05$&$0.563$&$2.9$\\
$g^2_{\pi\pi\sigma}/4\pi$&$0.6$&$0.6$&$0.25$\\
\hline\end{tabular}
\end{center}
\end{table}

\newpage
\begin{center}
{\large\bf FIGURE CAPTIONS}
\end{center}
     
Fig.1 One- and two-body contributions to the bremsstrahlung current
amplitude.
\vspace{0.5cm}

Fig.2  (a) Gauge-invariant bremsstrahlung current amplitude. (b) Post-,
pre-  and double-scattering amplitude contributions (se Eq. (9)). 
\vspace{0.5cm}

Fig.3 Born amplitude corresponding to the $\pi N$ potential
operator $\Uop$. The first diagram denote the nucleon-pole, and
the $\Delta$-pole corresponds to the third diagram.
 The fifth and sixth diagrams correspond to $\rho$ and $\sigma$ mesons
exchange.
\vspace{0.5cm}

Fig.4 The gauge-invariant amplitude obtained by coupling a photon to the
Born terms in fig.3, together with the two-meson exchange currents
(fifth, sixth and  seventh graphs in each line): (a) Contributions
obtained from the N and $\Delta$ direct-pole diagrams in
fig.3; (b) Terms generated by the cross-pole diagrams in fig.3, and (c)
diagrams obtained from $\rho$ and $\sigma$ exchange contributions in
fig.3.
\vspace{0.5cm}
 
Fig.5 $\piNg$ cross section for $T_{lab} = 269$, $298$ and
$324 MeV$ and  $G_{14} \equiv (\theta_\gamma = 103^0, \phi_\gamma =
180^0)$ in EXPI, calculated in the DMA for the different
T-matrices. We also include  the SPA cross section and the measured
values.
\vspace{0.5cm}

Fig.6 Same as fig.5  for $T_{lab} = 299 MeV$. The cross sections are averaged over the angles
$\theta_\pi =55-75^0$ (upper plot) and $\theta_\pi =75-95^0$(lower plot) in EXPII.

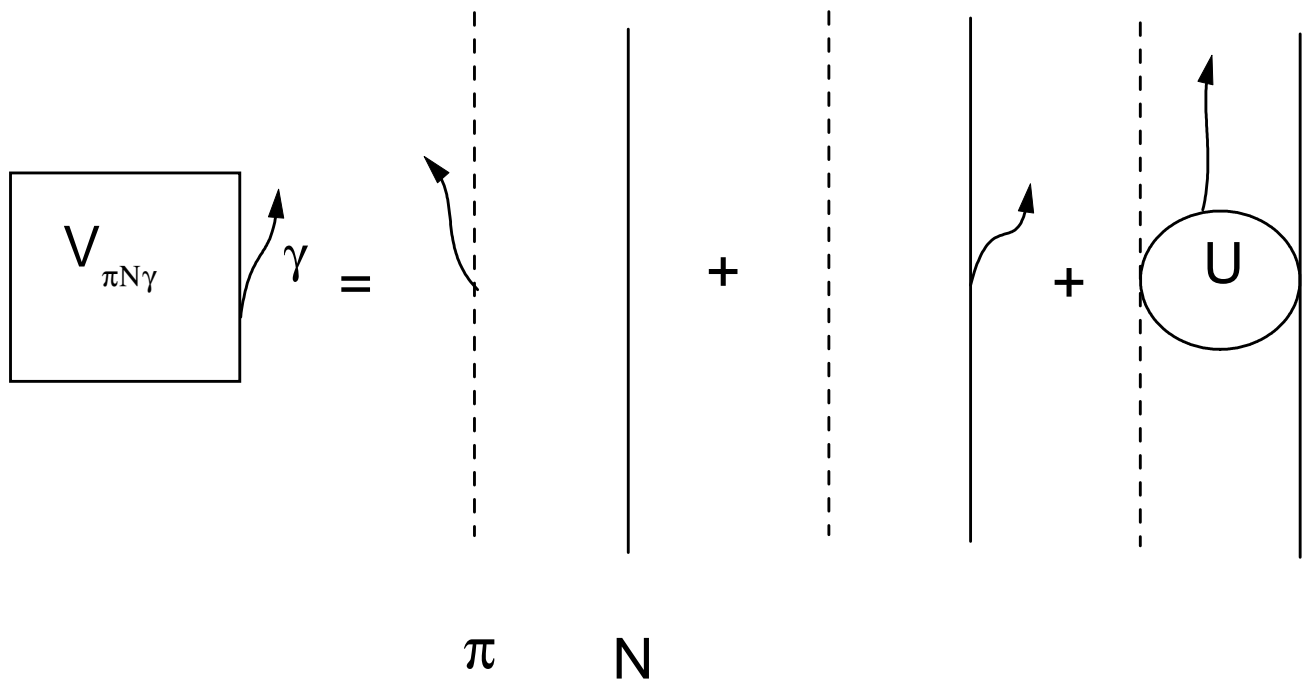
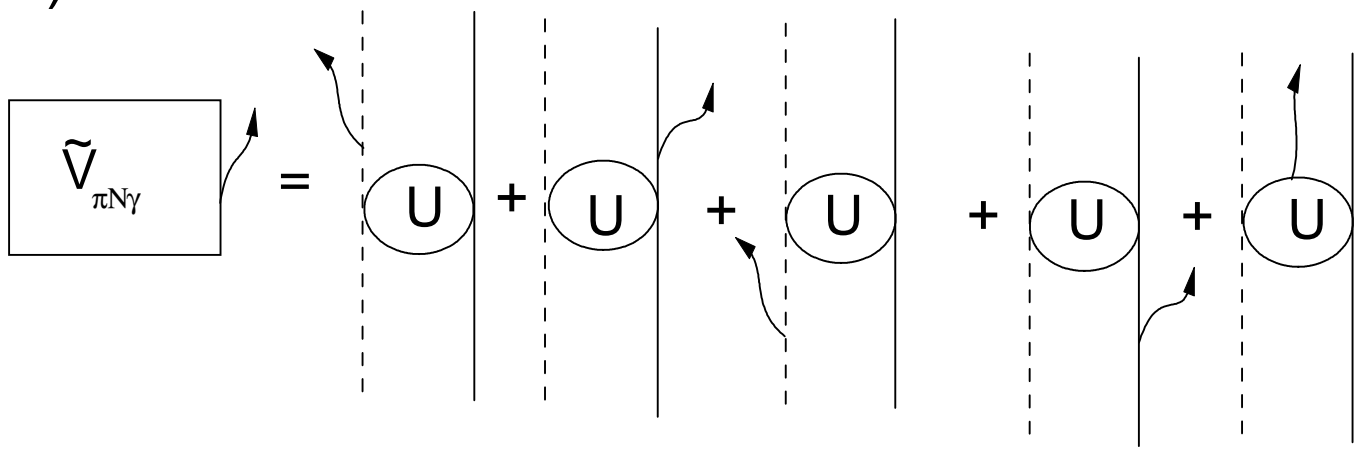
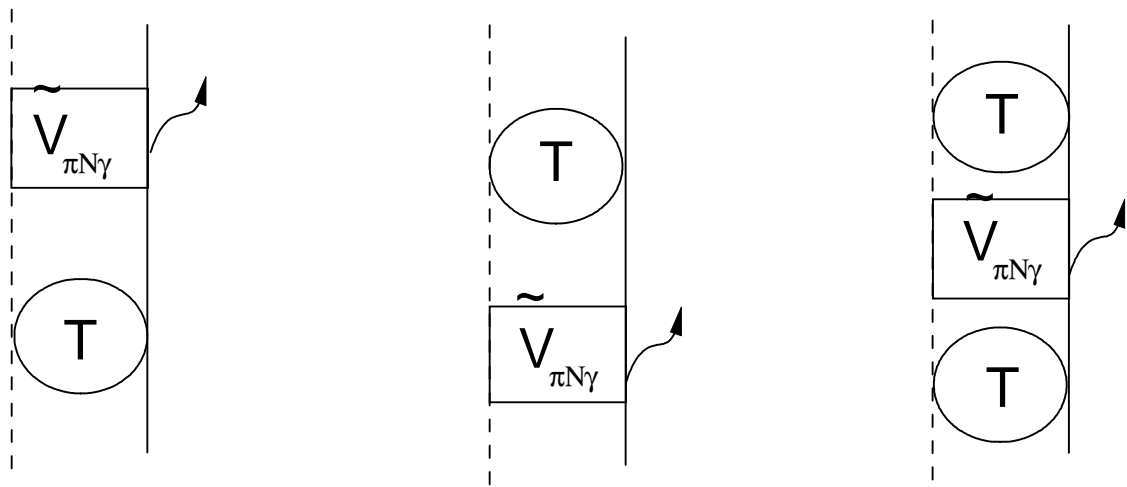
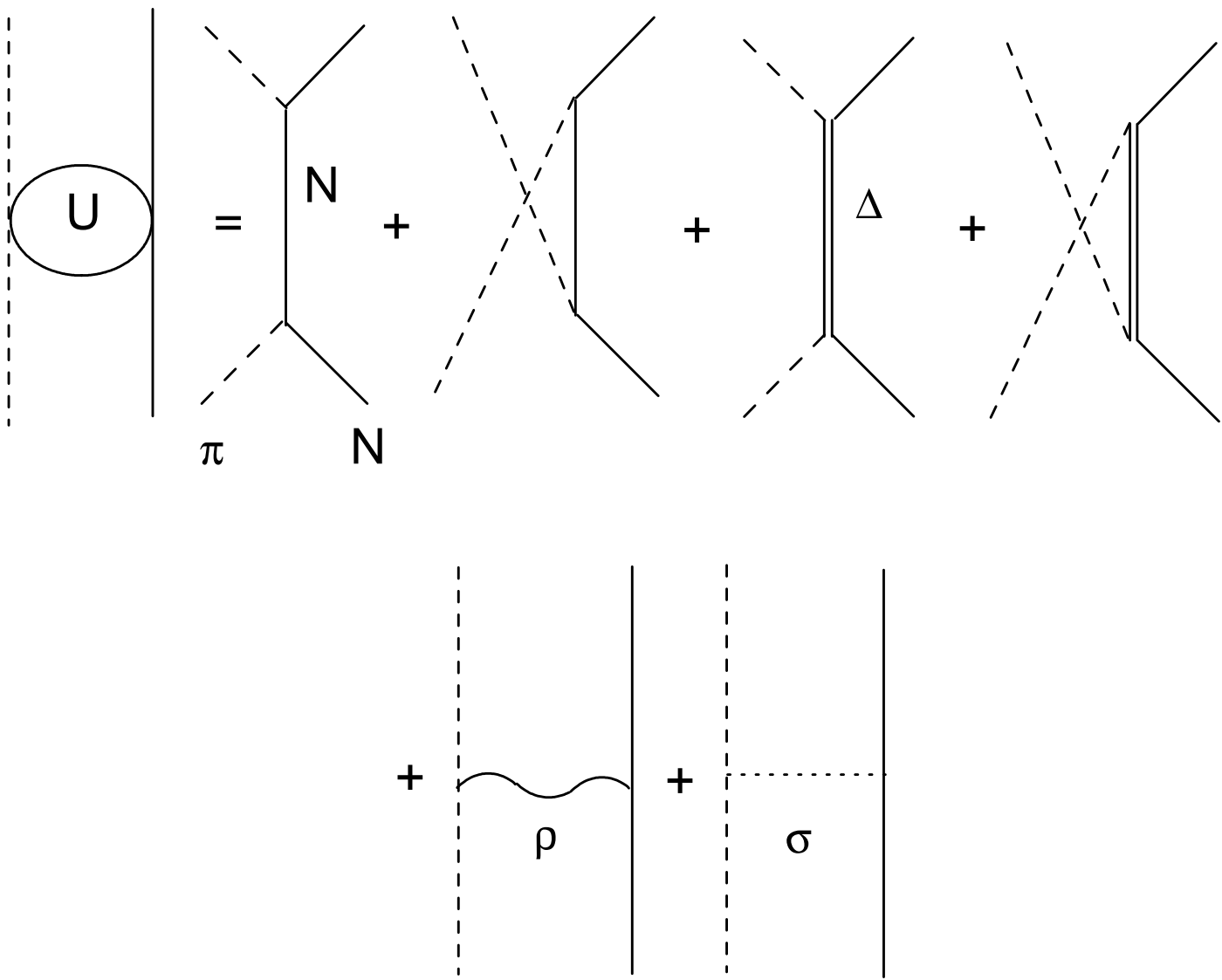
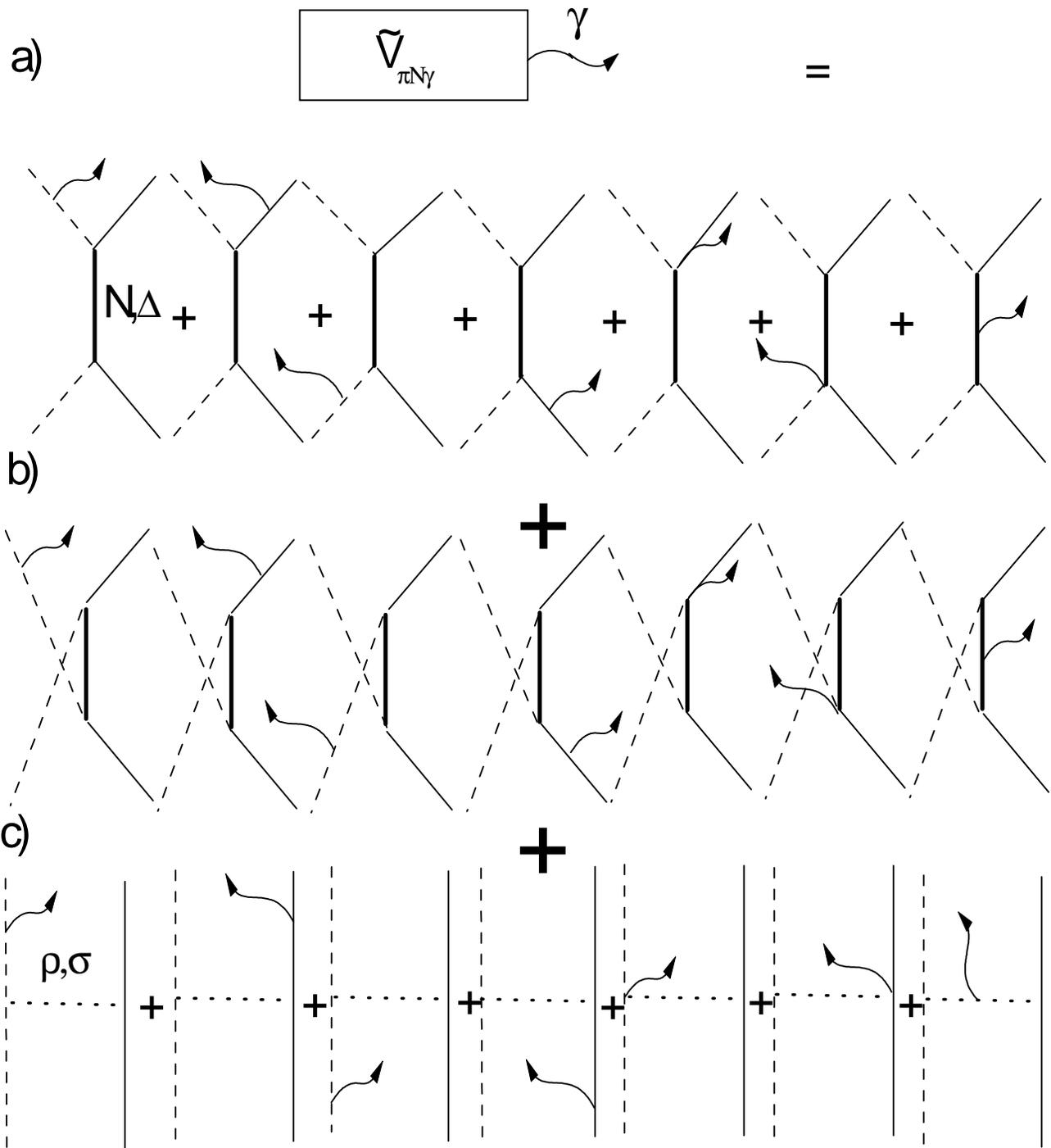
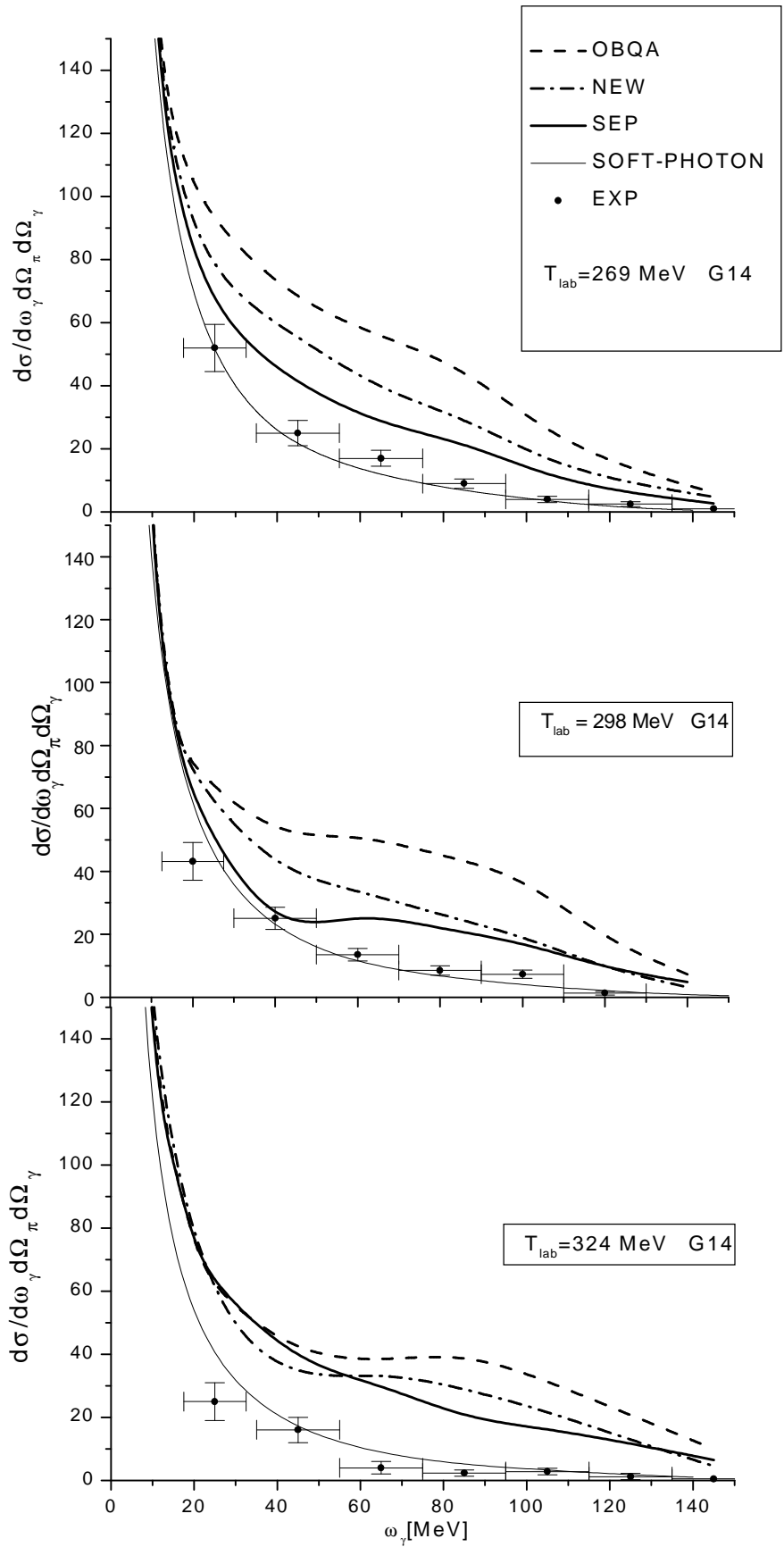
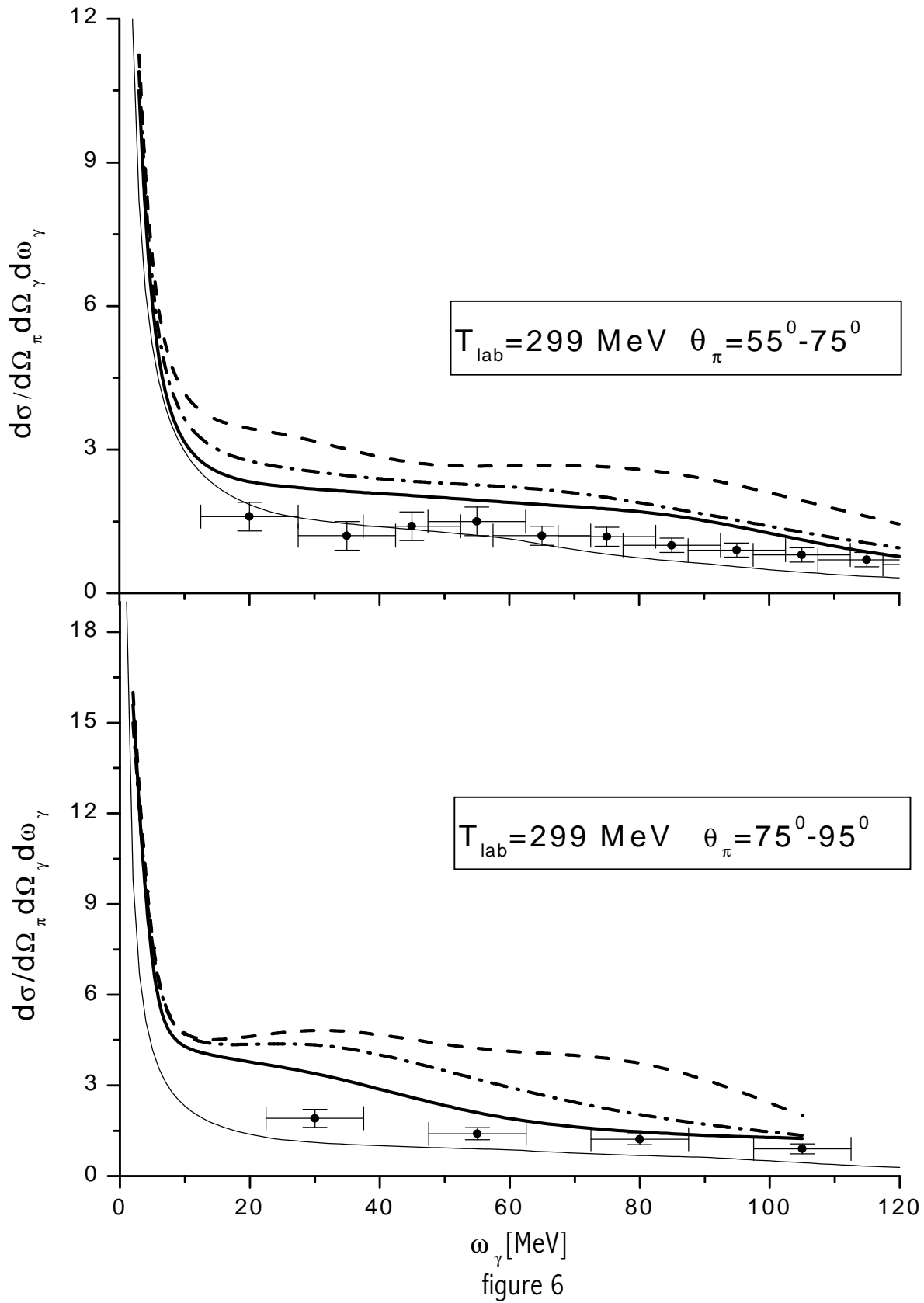
\end{document}